\documentstyle[12pt]{article}
\begin{document}
\begin {center}
{\bf {\Large  Low energy $\omega (\to \pi^0\gamma)$ meson photoproduction
              in the Nucleus} }
\end {center}
\begin {center}
Swapan Das \footnote {email: swapand@barc.gov.in} \\
{\it Nuclear Physics Division,
Bhabha Atomic Research Centre  \\
Mumbai-400085, India }
\end {center}

\begin {abstract}
The $\pi^0 \gamma$ invariant mass distribution spectra in the
$ (\gamma, \pi^0\gamma) $ reaction were measured by TAPS/ELSA collaboration
to look for the hadron parameters of the $\omega$ meson in Nb nucleus.
We study the mechanism for this reaction, where we consider that the
elementary reaction in Nb nucleus proceeds as
$ \gamma N \to \omega N; ~ \omega \to \pi^0\gamma $.
The $\omega$ meson photoproduction amplitude for this reaction is extracted
from the measured four momentum transfer distribution in the $ \gamma p \to
\omega p $ reaction. The propagation of the $\omega$ meson and the distorted
wave function for the $\pi^0$ meson in the final state are described by
the eikonal form.
The $\omega$ and $\pi^0$ mesons nucleus optical potentials, appearing
in the $\omega$ meson propagator and $\pi^0$ meson distorted wave function
respectively, are estimated using the ``$t\varrho$'' approximation.
The effects of pair correlation and color transparency are also studied.
The calculated results do not show  medium modification for the $\omega$
meson produced in the nucleus for its momentum greater than 200 MeV. It
occurs since the $\omega$ meson dominantly decays outside the nucleus.
The
dependence of the cross section on the final state interaction is also
investigated. The broadening of the $\omega$ meson mass distribution
spectra is shown to occur due to the large resolution width associated with
the detector used in the experiment.
\end {abstract}


PACS number(s): 25.20.Lj, 13.75.Lb

\section{Introduction}

The study for the properties of vector meson in the nuclear medium is a
topic of intense interest in the nuclear physics. This topic has been
pursued vigorously in recent years to determine quantitatively the mass
and width of the vector meson in the nucleus.
Various model calculations relate the in-medium properties of vector meson
with the chiral symmetry in QCD. The restoration of this symmetry in the
nuclear medium predicts the reduction of vector meson mass which could be
drastic in the hot and/or dense nuclear medium \cite{redv}.
Using the scaling property in QCD, Brow and Rho \cite{br} have shown that
the vector meson mass should drop in the nuclear medium. Other model
calculations, such as QCD sum rule calculations due to Hatsuda et al.
\cite{hat}, VMD model calculations due to Asakawa et al. \cite{aks}, and the
quark meson coupling model calculation due to Saito et al. \cite{sai},
corroborate this finding. On the other hand, the vector meson mass shift
(upward) and width broadening in the hot and dense nuclear matter are also
reported
\cite{ubrw}.

Large medium modification of the $\rho$ meson was indicated first in the
enhanced dilepton yield (between 300 and 700 MeV) in CERES and HELIOS
ultra-relativistic heavy ion collision data taken sometime around
1995 in CERN-SPS \cite{dilpex}. The quantitative estimation of this
modification could not be made in that time because of the poor statistics
and resolution of the data. Theoretically, these data have been found
compatible with both scenarios: (i) the dropping of $\rho$ meson mass
\cite{rhth1}, and (ii) the many body interaction of $\rho$ meson with other
hadrons in the nuclear medium \cite{rhth2}.
Almost after a decade, the STAR experiment at RHIC BNL \cite{rhic} found
the decrease in $\rho$ meson mass $\sim 70$ MeV in the analysis of the
$\pi^+\pi^-$ production data from the peripheral Au+Au collisions. However,
the upgraded CERES experiment \cite{uceres} as well as the dimuon
measurements (in the In-In collision) in the NA60 experiment at CERN
\cite{na60} reported considerable broadening in the $\rho$ meson mass
distribution spectrum, but essentially no shift in mass.

The interpretation of the ultra-relativistic heavy-ion collision data is
very complicated because this reaction occurs far from the equilibrium
state whereas the calculations for the hadronic parameters are done
at normal nuclear density and zero temperature (equilibrium state).
Therefore, it is more accountable to search the in-medium properties of
vector meson in the normal nucleus.
In fact, the modification of the vector meson is predicted large enough to
observe it in the nuclear reaction with pion and photon beams \cite{effn}.
The chiral symmetry is also shown to restore partially in the normal nucleus
(specifically, inside a heavy nucleus \cite{birse}). To be added, the
scaling hypothesis \cite{br} and QCD sum rule calculation \cite{hat}
envisage the reduction $(\sim 15-20\%)$ of vector meson mass in a
nucleus.
There also exist many body calculations which show the drop of mass and
increase in width for the vector meson in the normal nucleus
\cite{effn, mwnc}. Beside these, only width broadening \cite{ownc, herap},
upward mass-shift \cite{umnc, cov} and appearance of additional peaks
\cite{herap, cov, mpnc} for the vector meson in the nuclear medium are
predicted by various model calculations.

The modification of the vector meson in the normal nucleus has been
reported by various measurements. The KEK-PS E325 collaboration at KEK
\cite{kek} found the enhancement in the $e^+e^-$ yield in the $p+$A reaction
at 12 GeV. This enhancement is well understood due to the reduction of the
vector meson mass in the nucleus. The sub-threshold $\rho$ meson production
experiment on the $ ^3 \mbox{He} (\gamma, \pi^+\pi^-) \mbox{X} $ reaction,
done by the TAGX collaboration \cite{lolo}, reported large decrease in the
$\rho$ meson mass. The $\rho$ meson polarization distribution, measured
by this collaboration \cite{huber}, corroborates this finding. The
recent results, published from Jefferson Laboratory, show the broadening of
the $\rho$ meson width in the photonuclear reaction \cite{tjnaf}. The mass
shift of the $\rho$ meson found in this experiment is insignificant.

Recent past, the CBELSA/TAPS collaboration measured the $\pi^0\gamma$
invariant mass distribution spectrum to look for the medium modification
of the $\omega$ meson in Nb nucleus \cite{elsa}. This experiment was
done at the electron stretcher accelerator (ELSA) in Bonn using the tagged
$\gamma$ beam of energy spread $0.64-2.53$ GeV. The $\omega$ meson was
detected through its decay $ \omega \to \pi^0\gamma $.
To minimize the pionic distortion on the $\omega$ meson signal, the data
were taken for the $\pi^0$ meson kinetic energy $T_{\pi^0}$ larger
than 150 MeV.
The measured spectrum for the lower $\omega$ meson momentum $k_\omega$
bin (i.e., $ 200 < k_\omega (\mbox{MeV/c}) < 400 $) shows distinct $\omega$
meson mass modification at $\sim 730$ MeV, which gradually vanishes with
the increase in the $\omega$ meson momentum.
However, this claim is no more valid \cite{metag} as the reanalysis of the
data could not reproduced the spectral shape reported in Ref.~\cite{elsa}.
The measured distributions show broad width ($\sim$ 55 MeV) comparable to
the resolution width of the detecting system used in the experimental
set-up. In fact, this width is about 6.5 times larger than the free space
decay width of the $\omega$ meson ($ \Gamma_\omega = 8.43 $ MeV
\cite{pdg}). Therefore, the width of the $\omega$ meson in a nucleus
$\Gamma^*_\omega$ can be detected in this set-up if it is enhanced more
than 55 MeV. The present status of this topic is summarized in the recent
review articles \cite{rart}.

We study microscopically the mechanism for the $(\gamma,\pi^0\gamma)$
reaction on a nucleus, and compare our results with the data taken by the
CBELSA/TAPS collaboration. In the energy region of this reaction, the
elementary reaction occurring in the nucleus can be visualized as
$ \gamma N \to V N ; ~ V \to \pi^0\gamma $, where $V$ stands for the low
energy vector mesons, i.e., $\rho^0, ~ \omega, ~ \phi$ mesons.
The qualitative analysis presented in Ref.~\cite{das1} shows that the
$\pi^0\gamma$ event in the elementary reaction arises distinctly due to
the decay of the $\omega$ meson produced in the intermediate state. The
contributions to this event from the $\rho^0$ and $\phi$ mesons are found
negligible.
Therefore, we consider that the $(\gamma,\pi^0\gamma)$ reaction on the
nucleus proceeds through the formation of $\omega$ meson in the intermediate
state. The production of this meson is described by the
$\gamma N \to \omega N$ reaction amplitude $ f_{\gamma N \to \omega N} $,
and its propagation is presented by the eikonal form. The $\omega$ meson
interaction with the nucleus (which appears in its propagator) is described
by the corresponding optical potential. The decay of $\omega$ meson to
$\pi^0$ and $\gamma$ bosons is governed by the $\omega\pi\gamma$ Lagrangian.
The $\pi^0$ meson scattering state is generated by using the $\pi^0$ meson
nucleus optical potential.

\section{Formalism}

The generalized optical potential or self-energy for the photoproduction
of $\omega$ meson in a nucleus \cite{das1, pash} is given by
\begin{equation}
\Pi_{\gamma A \to \omega A} ( {\bf r} ) =
-4\pi \left [ 1 + \frac{E_\omega}{E_N} \right ]
f_{\gamma N \to \omega N} (0) \varrho ({\bf r}),
\label{gpa}
\end{equation}
where $\varrho ({\bf r})$ represents the spatial density distribution of
the nucleus.

The factor $ f_{\gamma N \to \omega N} (0) $ in Eq.~(\ref{gpa}) is the
forward reaction amplitude for the elementary $\gamma N \to \omega N$
process. It is related to the four momentum $q^2$ transfer distribution
$ d\sigma (\gamma N \to \omega N) / dq^2 $ \cite{shmt}:
\begin{equation}
\frac{d\sigma}{dq^2} ( \gamma N \to \omega N ; q^2=0 )
=\frac{\pi}{k^2_\gamma} | f_{\gamma N \to \omega N} (0) |^2.
\label{fgo2}
\end{equation}
The forward $ d\sigma (\gamma p \to \omega p) / dq^2 $ is used to be obtained
by extrapolating the measured value of
$ d\sigma (\gamma p \to \omega p; q^2) / dq^2 $ at $q^2=0$. In fact, the
energy dependent values for it are reported in Refs.~\cite{shmt, stk} for
$ E_\gamma \ge 1.6 $ GeV. At lower energies, i.e., $ E_\gamma \le 2.6 $
GeV, the four-momentum transfer distributions
$ d\sigma (\gamma p \to \omega p) / dq^2 $ were measured with  SAPHIR
detector at electron stretcher ring (ELSA), Bonn \cite{bar1}. We extract the
energy dependent $ | f_{\gamma p \to \omega p} (0)|^2 $ from the measured
$ d\sigma (\gamma p \to \omega p) / dq^2 $, and use them in our calculation.

The omega meson, produced in the nucleus, propagates certain distance
before it decays into $\pi^0$ and $\gamma$ bosons. The propagation of the
$\omega$ meson from its production point ${\bf r}$ to its decay point
${\bf r^\prime}$ can be expressed as
$ ( -g^\mu_{\mu^\prime} + \frac{1}{m^2} k^\mu_\omega k_{\omega,\mu^\prime} )
G_\omega ( m; {\bf r^\prime - r} ) $ \cite{das3}. We represent the scalar
part of the $\omega$ meson propagator
$ G_\omega ( m; {\bf r^\prime - r} ) $
by the eikonal form \cite{das5, gkc}, i.e.,
\begin{equation}
G_\omega ( m; {\bf r^\prime - r} ) = \delta ( {\bf b^\prime - b} )
\Theta (z^\prime - z) e^{ i {\bf k_\omega . (r^\prime - r)} }
D_{\bf k_\omega} ( m; {\bf b}, z^\prime, z ).
\label{omp}
\end {equation}
The factor $ D_{\bf k_\omega} ( m; {\bf b}, z^\prime, z ) $ appearing in this
equation describes the nuclear medium effect on the properties of $\omega$
meson. The form for it is
\begin{equation}
D_{\bf k_\omega} ( m; {\bf b}, z^\prime, z ) =
-\frac{i}{ 2k_{\omega\parallel} }
exp \left [  \frac{i}{ 2k_{\omega\parallel} } \int_z^{z^\prime}
 dz^{\prime \prime} \{ {\tilde G}^{-1}_{0\omega} ( m ) -
    2 E_\omega V_{O\omega} ({\bf b}, z^{\prime \prime}) \} \right  ],
\label{dom}
\end{equation}
where $k_\omega$ is the momentum of the $\omega$ meson.
$ V_{O\omega} ({\bf b}, z^{\prime \prime}) $ represents the $\omega$ meson
nucleus optical potential which arises due to the interaction of this
meson with the particles present in the nucleus. In fact, this potential
modifies the hadronic parameters of the $\omega$ meson during its passage
through the nucleus.
$ {\tilde G}_{0\omega} (m) $ denotes the $\omega$ meson (on-shell)
propagator in free space: $ {\tilde G}^{-1}_{0\omega} (m)
= m^2-m^2_\omega+im_\omega\Gamma_\omega (m) $. Here, $m_\omega$ and
$\Gamma_\omega (m)$ represent the resonant mass and total decay width for
the $\omega$ meson, elaborated in Ref.~\cite{das1}.

The $\pi^0\gamma$ in the final state originates due to the decay of
$\omega$ meson, i.e., $ \omega \to \pi^0\gamma $. The Lagrangian density
$ {\cal L} _ {\omega\pi\gamma} $ describing this decay channel
\cite{das1, lcsr} is given by
\begin{equation}
{\cal L}_{\omega\pi\gamma} = - \frac{ f_{\omega\pi\gamma} }{ m_\pi }
\epsilon_{\mu\nu\rho\sigma} \partial^\mu A^\nu \pi
\partial^\rho \omega^\sigma,
\label{lag}
\end{equation}
where $ f_{\omega\pi\gamma} (=0.095) $ denotes the $\omega\pi\gamma$
coupling constant. $A^\nu$ represents the photon field.

The wave function for photon is described by the plane wave. The distorted
wave function $ \chi^{(-)} ( {\bf k_{\pi^0}, r^\prime } ) $ for the
$\pi^0$ meson in the final state is represented by the eikonal form
\cite{das5, glub}:
\begin{equation}
\chi^{(-)*} ( {\bf k_{\pi^0}, r^\prime } )
= e^{ -i {\bf k_{\pi^0}.r^\prime} }
  D^{(-)*}_{\bf k_{\pi^0}} ({\bf b}, z^\prime).
\label{dwpi1}
\end{equation}
The factor $ D^{(-)*}_{\bf k_{\pi^0}} ({\bf b}, z^\prime) $ in this equation
describes the pionic distortion. The form for it is
\begin{equation}
D^{(-)*}_{\bf k_{\pi^0}} ({\bf b}, z^\prime)
= exp \left [ - \frac{i}{v_{\pi^0\parallel}} \int^\infty_{z^\prime}
dz_1  V_{O\pi^0} ({\bf b}, z_1) \right ],
\label{dwpi2}
\end{equation}
where $v_{\pi^0}$ is the velocity of pion.
$ V_{O\pi^0} ({\bf b}, z_1) $ denotes the optical
potential for the pion nucleus scattering in the final state.

The differential cross section for the $\omega$ meson mass $m$ (i.e.,
the $\pi^0 \gamma$ invariant mass) distribution can be written as
\begin{equation}
\frac{ d\sigma (m,E_\gamma) }{ dm }
= \int d\Omega_\omega K_F \Gamma_{\omega \to \pi^0\gamma^\prime} (m)
                   | F({\bf k}_\gamma, {\bf k}_\omega) |^2,
\label{dsc1}
\end{equation}
where $K_F$ is the kinematical factor of the reaction. It is given by
$ K_F = \frac{1}{(2\pi)^3} \frac{k_\omega}{k_\gamma} m^2 $; with
$ {\bf k}_\omega = {\bf k}_{\pi^0} + {\bf k}_{\gamma^\prime} $. The prime
represents the quantity in the final state.

$ \Gamma_{\omega \to \pi^0\gamma} (m) $ in above equation denotes the
width for the $\omega$ meson of mass $m$ decaying at rest into
$\pi^0\gamma$ channel. $ \Gamma_{\omega \to \pi^0\gamma} (m) $ is evaluated
\cite{das1} using the Lagrangian density $ {\cal L_{\omega\pi\gamma}} $
given in Eq.~(\ref{lag}):
\begin{equation}
\Gamma_{\omega \to \pi^0\gamma} (m)
=\Gamma_{\omega \to \pi^0\gamma} (m_\omega)
 \left [ \frac{k(m)}{k(m_\omega)} \right ]^3,
\label{wdth2}
\end{equation}
with $ \Gamma_{\omega \to \pi^0\gamma} (m_\omega=782 ~\mbox{MeV}) \approx
0.72$ MeV. $k(m)$ is the momentum of pion in the $\pi^0\gamma$ cm system.

The factor $ F ({\bf k}_\gamma, {\bf k}_\omega) $ in Eq.~(\ref{dsc1})
describes the production of $\omega$ meson in the nucleus. In addition,
it also carries the information about the $\omega$ meson propagation inside
as well as outside the nucleus. The expression for it is
\begin{equation}
F ({\bf k}_\gamma, {\bf k}_\omega)
= \int d{\bf r} \Pi_{\gamma A \to \omega A} ({\bf r})
   e^{ i({\bf k}_\gamma - {\bf k}_\omega) . {\bf r} }
   D ({\bf k}_\omega; {\bf b}, z),
\label{fdfn}
\end{equation}
where $ D ({\bf k}_\omega; {\bf b}, z) $ is given by
\begin{equation}
D ({\bf k}_\omega; {\bf b}, z)
= \int^\infty_z dz^\prime  D^{(-)*}_{\bf k_{\pi^0}} ({\bf b}, z^\prime)
   D_{\bf k_\omega} ({\bf b}, z^\prime, z).
\label{ybio}
\end{equation}
All quantities appearing in this equation are already defined. The $\omega$
meson decay probabilities inside and outside the nucleus can be addressed
by splitting $ D ({\bf k}_\omega; {\bf b}, z) $ in Eq.~(\ref{ybio}) into
two parts:
\begin{equation}
D ({\bf k}_\omega; {\bf b}, z)
= D_{in} ({\bf k}_\omega; {\bf b}, z) + D_{out} ({\bf k}_\omega; {\bf b}, z),
\label{ybio2}
\end{equation}
where $ D_{in} ({\bf k}_\omega; {\bf b}, z) $ and
$ D_{out} ({\bf k}_\omega; {\bf b}, z) $ describe the $\omega$ meson
decay inside and outside the nucleus respectively. Using Eq.~(\ref{ybio}),
they can be written as
\begin{eqnarray}
&D_{in} ({\bf k}_\omega; {\bf b}, z)&
= \int^Z_z dz^\prime  D^{(-)*}_{\bf k_{\pi^0}} ({\bf b}, z^\prime)
   D_{\bf k_\omega} ({\bf b}, z^\prime, z),    \\
\label{ybi}
&D_{out} ({\bf k}_\omega; {\bf b}, z)&
= \int^\infty_Z dz^\prime  D^{(-)*}_{\bf k_{\pi^0}} ({\bf b}, z^\prime)
   D_{\bf k_\omega} ({\bf b}, z^\prime, z),
\label{ybo}
\end{eqnarray}
with $ Z=\sqrt{R^2-b^2} $. $R$ is the extension of the nucleus.
In fact, $ D_{out} ({\bf k}_\omega; {\bf b}, z) $ in the above equation can
be simplified to $ {\tilde G}_{0\omega} (m) exp [ \frac{i}{2k_\omega}
{\tilde G}^{-1}_{0\omega} (m) (Z-z) ] $.

The Eq.~(\ref{dsc1}) illustrates the differential cross section for the
$\omega$ meson mass distribution due to fixed $\gamma$ beam energy
$E_\gamma$. But, as we mentioned earlier, the CBELSA/TAPS collaboration
\cite{elsa} used tagged photon beam of definite energy range in their
measurement. Therefore, we modulate the cross section in Eq.~(\ref{dsc1})
with the beam profile function $W(E_\gamma)$ \cite{kho}, i.e.,
\begin{equation}
\frac{d\sigma (m)}{dm} = \int^{E_\gamma^{mx}}_{E_\gamma^{mn}}
dE_\gamma W(E_\gamma) \frac{d\sigma (m,E_\gamma)} {dm}.
\label{dsc2}
\end{equation}
$E_\gamma^{mn}$ and $E_\gamma^{mx}$ are equal to 0.64 GeV and 2.53 GeV
respectively, as provided in Ref.~\cite{elsa}. The profile function
$W(E_\gamma)$ for the $\gamma$ beam (originating due to the bremsstrahlung
radiation of electron impinging on the Pb target \cite{elsa}) varies as
$ W(E_\gamma) \propto \frac{1}{E_\gamma} $ \cite{kho}.

\section{Results and Discussions}

We calculate the $\omega$ meson mass $m$ (i.e., the $\pi^0 \gamma$ invariant
mass) distribution spectra in the $ ( \gamma, \omega \to \pi^0 \gamma ) $
reaction on $^{93}$Nb nucleus. The $\omega$ meson, as mentioned earlier,
is detected for pion (arising due to $\omega \to \pi^0\gamma$) kinetic
energy $T_{\pi^0}$ greater than 150 MeV. We impose this condition in our
calculation, i.e., all results presented in this manuscript are calculated
for $T_{\pi^0} > 150$ MeV.

The meson nucleus optical potentials, i.e., $V_{O\omega}$ in Eq.~(\ref{dom})
and $V_{O\pi^0}$ in Eq~(\ref{dwpi2}), are estimated using the
$``t\varrho"$ approximation \cite{das5}:
\begin{equation}
V_{OM} ({\bf r})
= - \frac{v_M}{2} [i+\alpha_{MN}] \sigma_t^{MN} \varrho ({\bf r}).
\label{opt}
\end{equation}
The symbol $M$ in this equation stands for a meson (i.e., either $\omega$
meson or $\pi^0$ meson), and $N$ denotes a nucleon. $v_M$ is the velocity
of the meson $M$.
$\alpha_{MN}$ represents the ratio of the real to imaginary part of the
meson nucleon scattering amplitude $f_{MN}$, and $\sigma_t^{MN}$ is the
corresponding total cross section.
Since the $\omega$ meson is a neutral unstable particle, $\alpha_{\omega N}$
and $\sigma_t^{\omega N}$ can't be obtained directly from measurements.
Lutz et al., \cite{lwf} (using couple channel calculation) have estimated
the energy dependent  $f_{\omega N}$ in the low energy region, i.e.,
$ 1.4 \le \sqrt{s} \mbox{(GeV)} \le 1.8 $. Their calculations are well
constrained by the elementary $\omega$ meson production data in the
threshold region.
At higher energy, the imaginary part of $f_{\omega N}$ is extracted from the
elementary $\omega$ meson photoproduction data using vector meson dominance
model \cite{lcsr, ses}. Sibirtsev et al. \cite{ses}, using additive quark
model and Regge theory, have evaluated $f_{\omega N}$ for a wide range of
energy. According to them $\alpha_{\omega N}$ can be written as
$ \alpha_{\omega N}
= \frac{ 0.173(s/s_0)^\epsilon - 2.726(s/s_0)^{-\eta} }
       { 1.359(s/s_0)^\epsilon + 3.164(s/s_0)^{-\eta} } $,
with $ s_0=1 ~\mbox{GeV}^2 $, $ \epsilon=0.08 $ and $ \eta=0.45 $ \cite{ses}.
For $V_{O\pi^0} ({\bf r})$ also, the energy dependent $\alpha_{\pi^0 N}$
and $\sigma_t^{\pi^0 N}$ can't be measured directly. Therefore, we have
worked out the $\pi^0$ meson nucleon scattering amplitude $f_{\pi^0N}$ using
isospin algebra:
$ f_{\pi^0N} = \frac{1}{2} [ f_{\pi^+N} + f_{\pi^-N} ] $. The energy
dependent measured values for the $\pi^\pm N$ scattering parameters, i.e.,
$\alpha_{\pi^\pm N}$ and $\sigma_t^{\pi^\pm N}$, are available in
Ref.~\cite{pdg2}.

The nuclear density distribution $ \varrho ({\bf r}) $, required to evaluate
Eqs.~(\ref{gpa}) and (\ref{opt}), is approximated by the nuclear charge
density distribution. The form of $ \varrho ({\bf r}) $ for $^{93}$Nb
nucleus, as obtained from the electron scattering data, is given by
\begin{equation}
\varrho ({\bf r}) = \varrho_0 \frac{ 1 }{ 1+exp(\frac{r-c}{z}) };
\label{dnb}
\end{equation}
with c=4.87 fm, z=0.573 fm \cite{andt}. It is normalized to the mass number
of the nucleus.
The pair correlation (PC) in $ \varrho ({\bf r}) $ can be incorporated
by the following replacement \cite{pash}:
\begin{equation}
\varrho ({\bf r}) \to \varrho ({\bf r}) \left [ 1 + \frac{1}{2}
\sigma^{MN}_t \varrho ({\bf r})  l_c
\left \{ \frac{ \varrho({\bf r}) }{ \varrho(0) } \right \}^{1/3}
\right ] ,
\label{prcn}
\end{equation}
where the correlation length $l_c$ is equal to 0.3 fm \cite{pash}. We
include $\alpha_{MN}$ in the above equation by the trivial substitution
$ \sigma^{MN}_t \to (1-i\alpha_{MN}) \sigma^{MN}_t $ \cite{glub}.

The onset of color transparency (CT) has been reported in many nuclear pion
production reactions at intermediate energies \cite{ct1, ct2}. The color
transparency arises due to the reduction of the elementary pion nucleon
total cross section over a typical length scale, commonly known as hadron
formation length $l_h$. The standard expression for the effective cross
section (which incorporates CT effect in it \cite{ct2}) is
\begin{equation}
\frac{ \sigma^{\pi^0N}_{t, eff} }{ \sigma^{\pi^0N}_t } =
\left [ \left \{ \frac{z}{l_h} +
\frac{n^2k^2_t}{Q^2} \left ( 1 - \frac{z}{l_h} \right ) \right \}
\theta (l_h-z) + \theta (z-l_h) \right ].
\label{ect}
\end{equation}
The value of $n$ is equal to 2 for a quark-antiquark color singlet system.
$k_t$ (=0.35 GeV/c) is the average transverse momentum of a quark inside the
hadron. $z$ is the straight path travelled by the pion after its formation.
In above equation, $ Q^2 = | (k^\mu_\gamma - k^\mu_{\gamma^\prime})^2 | $,
and $ l_h \equiv l_{\pi^0} \simeq 2 k_{\pi^0} / ( 0.7 ~ \mbox{GeV}^2 ) $
\cite{ct2}.

The main purpose of this calculation is to look for the in-medium properties
of $\omega$ meson in the nucleus. We, therefore, present in Fig.~1
the $\omega$ meson mass distribution spectra calculated with and without
the $\omega$ meson nucleus (i.e., Nb nucleus) interaction for
$ k_\omega \mbox{(GeV/c)} = 0.21-0.39 $. As mentioned earlier, it is the
lowest $\omega$ meson momentum bin in the measurement where the medium
modification was shown to occur \cite{elsa}.
The solid curve in Fig.~1 represents the previous case, i.e., the $\omega$
meson nucleus interaction is incorporated in the calculated cross section.
The cross section calculated without this interaction is presented by the
dot-dot-dashed curve. In both cases, the pion nucleus interaction has been
incorporated. This figure distinctly shows the absence of medium
modification for the $\omega$ meson photoproduced in the Nb nucleus
for $ 0.2 < k_\omega \mbox{(GeV/c)} < 0.4 $.

Since Fig.~1 does not show the medium modification of the $\omega$ meson
produced in the Nb nucleus, we envisage to compare the $\omega$ meson decay
probabilities inside and outside the nucleus for the same kinematical
condition quoted in Fig.~1, i.e., $ 0.2 < k_\omega \mbox{(GeV/c)} < 0.4 $.
In Fig.~2, we plot the mass distribution spectra for the $\omega$ meson
decaying inside and outside the Nb nucleus. This figure shows that the
$\omega$ meson decays dominantly outside the nucleus (dashed curve).
The cross section for the $\omega$ meson decaying inside the nucleus is
presented by the dot-dashed curve. The solid curve in Fig.~2 illustrates
that the coherent addition of the amplitudes of the $\omega$ meson decaying
inside and outside the nucleus enhances the cross section significantly at
the peak. For $ k_\omega > 0.4 $ GeV/c, the $\omega$ meson moves faster
resulting less interaction with the nucleus. In such case, the cross
section for the $\omega$ meson decaying outside the nucleus would be
relatively larger than that shown in Fig.~2. Therefore, the medium effect
on the $\omega$ meson in Nb nucleus is hardly possible under this
circumstances.

It should be mentioned that the $\omega$ meson decays throughout its
passage inside as well as outside the nucleus. The decay probability of
this meson inside the nucleus would be larger if its effective decay
length $ L^*_\omega (= v_\omega / \Gamma^*_\omega) $ in the nucleus is
less than the dimension of the nucleus. Here, $v_\omega$ and
$\Gamma^*_\omega$ denote the velocity and the effective width of the
$\omega$ meson in the nucleus.
Since the width of $\omega$ meson in the free state (i.e., $\Gamma_\omega$)
is equal to 8.43 MeV, the free space $\omega$ meson decay length
$ L_\omega (= v_\omega / \Gamma_\omega) $ is about 5.8 fm for $k_\omega$
equal to 200 MeV/c. This decay length would be much larger for $k_\omega$
equal to 400 MeV/c ($ L_\omega \sim 10.66 $ fm for $ k_\omega = 400 $
MeV/c).
Therefore, $L_\omega$ for $ k_\omega = 200 - 400 $ MeV is significantly
larger than the radius of $^{93}$Nb nucleus ($R_{rms} \sim 4.3$ fm). The
above analysis shows that the $\omega$ meson can't decay inside the Nb
nucleus unless its width in this nucleus is drastically increased, i.e.,
$ \Gamma^*_\omega >> \Gamma_\omega $.

We present in Fig.~3 the calculated results showing the effect of pionic
distortion on the $\omega$ meson mass distribution spectrum for
$ 0.2 < k_\omega \mbox{(GeV/c)} < 0.4 $. This figure shows that the
calculated cross section is not sensitive to the pion nucleus interaction.
The decay of $\omega$ meson $ (\to \pi^0\gamma) $ dominantly outside the
nucleus (shown in Fig.~2) causes the cross section insensitive to this
interaction.
The incorporation of color transparency (CT) in the pion nucleus
interaction (see Eq.~(\ref{ect})) does not make any change in the
calculated spectrum. Therefore, we do not show it. In fact, the cross
section, as mentioned above, is not sensitive to the pion nucleus
interaction itself.
It could be added that CT is a high energy phenomenon, i.e., $Q^2 \ge 1$.
Large hadron formation length $l_h$ (which occurs at higher pion momentum)
is also needed for the color transparency. Therefore, CT is not expected in
the low energy nuclear reaction, as considered in this manuscript.

The sensitivity of the pair correlation (PC) to the calculated cross
section is presented in Fig.~4. The solid curve in this figure shows the
calculated $\omega$ meson mass distribution spectrum where PC is included
in the nuclear density distribution
(see $\varrho ({\bf r})$ in Eqs.~(\ref{dnb}) and (\ref{prcn})).
The dashed curve represents the calculated spectrum without PC incorporated
in the nuclear density distribution, i.e., $\varrho ({\bf r})$ given in
Eq.~(\ref{dnb}). To be mentioned, PC is included in all other results
shown in this manuscript.
Fig.~4 shows that the incorporation of PC increases the magnitude of the
calculated cross section significantly, i.e., by a factor of $\sim 1.46$
at the peak.

The broad width appearing in the measured $\pi^0\gamma$ invariant mass
distribution spectra can be presumed due to the large resolution
width ($\sim$ 55 MeV) in the detecting system used in the experimental
set-up \cite{elsa}. We incorporate this in our formalism by folding
a Gaussian function $R(m,m^\prime)$ with the differential cross section
given in Eq.~(\ref{dsc2}), i.e.,
\begin{equation}
\frac{d\sigma (m)}{dm}
= \int dm^\prime R(m,m^\prime) \frac{d\sigma (m^\prime)}{dm^\prime}.
\label{dsc3}
\end{equation}
The function $R(m,m^\prime)$ in this equation accounts the resolution
width for the detector used in the measurement. It is given by
\begin{equation}
R(m,m^\prime)
= \frac{1}{\sigma\sqrt{2\pi}} e^{-\frac{(m-m^\prime)^2}{2\sigma^2}},
\label{resl}
\end{equation}
where $\sigma$ is related to the full width at half-maxima of the
resolution function $R(m,m^\prime)$: FWHM$=2.35\sigma$ \cite{knoll}. The
value of FWHM is taken equal to the width of the detector resolution,
i.e., 55 MeV as given in Ref.~\cite{elsa}. Fig.~5 shows that the
incorporation of the detector resolution function in the calculation
reduces the cross section at the peak, and it enhances simultaneously
the width of the cross section.

We present in Fig.~6 the calculated results due to Eq.~(\ref{dsc3}) along
with the measured $\pi^0\gamma$ invariant mass distribution spectra
(histograms, taken from the Ref.~\cite{elsa}) for the $\omega$ meson
momentum bins: (i) $0.6 < k_\omega \mbox{(GeV/c)} < 1$ and
(ii)  $1   < k_\omega \mbox{(GeV/c)} < 1.4$.
The solid curves appearing in this figure denote the calculated spectra
for  (i) $ k_\omega = 0.61 - 0.99 ~\mbox{GeV/c} $
and (ii)  $ k_\omega = 1.01 - 1.39 ~\mbox{GeV/c} $.
It is noticeable in this figure that the calculated distributions reproduce
the data very well. As mentioned earlier, the CB/TAPS collaboration
reported the modification of $\omega$ meson mass in the Nb nucleus in the
region below 780 MeV for
$0.2 < k_\omega \mbox{(GeV/c)} < 0.4$ and
$0.4 < k_\omega \mbox{(GeV/c)} < 0.6$ \cite{elsa}. Since this claim is no
longer valid \cite{metag}, we do not compare the calculated results with
the data for $ k_\omega < $ 0.6 GeV/c.

\section{Conclusions}

We have studied the mechanism for the $(\gamma, \pi^0\gamma)$ reaction on
Nb nucleus in the energy region: $0.64-2.54$ GeV. The $\pi^0$ and $\gamma$
bosons appearing in coincidence in the final state are shown to originate
due to the decay of $\omega$ meson produced in the intermediate state.
The
calculated results show that the properties of $\omega$ meson
in the nucleus are not modified for the kinematics used in the measurement
done at ELSA. It occurs since the $\omega$ meson dominantly decays outside
the nucleus.
This is also the reason for not showing the distortion due to final state
interaction, i.e., the $\pi^0$ meson nucleus interaction.
The
incorporation of color transparency in the pion nucleus interaction does not
change the cross section for the $\omega$ meson mass distribution.
The
inclusion of pair correlation in the nuclear density distribution increases
the magnitude of the cross section.
The
broad width in the measured distribution is found to arise due to the large
resolution width associated with the detector used in the experiment.
The calculated results reproduce the measured spectra very well.

\section{Acknowledgement}

I gratefully acknowledge Dr. L. M. Pant for making me aware about the
measurement for the omega meson mass distribution. The communication
made with Prof. Dr. E. Oset regarding the beam profile function is very
helpful. The discussion with Dr. D. R. Chakrabarty on the detector
resolution is highly appreciated.

\newpage

{\bf Figure Captions}
\begin{enumerate}

\item
(color online).
The calculated $\omega$ meson (photoproduced in $^{93}$Nb nucleus) mass
distribution spectra with (solid curve) and without (dot-dot-dashed curve)
the $\omega$ meson nucleus interaction for
$ k_\omega \mbox{(GeV/c)} = 0.21-0.39 $. The $\pi^0$ meson nucleus
interaction is incorporated in both cases.

\item
(color online).
The calculated mass distribution spectra for the $\omega$ meson decaying
inside as well as outside $^{93}$Nb nucleus. The dot-dashed curve represents
the cross section for the $\omega$ meson decaying inside the nucleus,
whereas the dashed curve corresponds to its decay outside the nucleus.
The coherent addition of them is shown by the solid curve.

\item
(color online).
The effect of pionic distortion on the $\omega$ meson mass distribution
spectrum for $ k_\omega \mbox{(GeV/c)} = 0.21-0.39 $.
The dotted curve represents the calculated spectrum where the plane wave
for $\pi^0$ meson is considered in the calculation. The incorporation of
the pionic distortion, as shown by the solid curve, hardly alter the plane
wave result.

\item
(color online).
The sensitivity of the pair correlation (PC) on the $\omega$ meson mass
distribution spectrum. The solid and dashed curves describe respectively
the calculated cross sections with and without PC incorporated in the
nuclear density distribution (see text).

\item
(color online).
The effect of the detector resolution on the $\omega$ meson mass
distribution spectrum. The solid curve corresponds to the cross section
given in Eq.~(\ref{dsc2}), where the detector resolution is not incorporated
in the calculation. The dot-dashed curve represents the cross section
where the detector resolution is incorporated in the calculation, see
Eq.~(\ref{dsc3}) in the text.

\item
(color online).
The calculated $\omega$ meson mass distribution spectra (solid curves due
to Eq.~(\ref{dsc3})) in the $(\gamma, \omega \to \pi^0 \gamma)$ reaction
on $^{93}$Nb nucleus for the $\omega$ meson momentum bins quoted
in the figure. The histograms represent the measured $\pi^0\gamma$ invariant
mass distribution spectra (taken from Ref.~\cite{elsa}), normalized to
the calculated cross sections.

\end{enumerate}

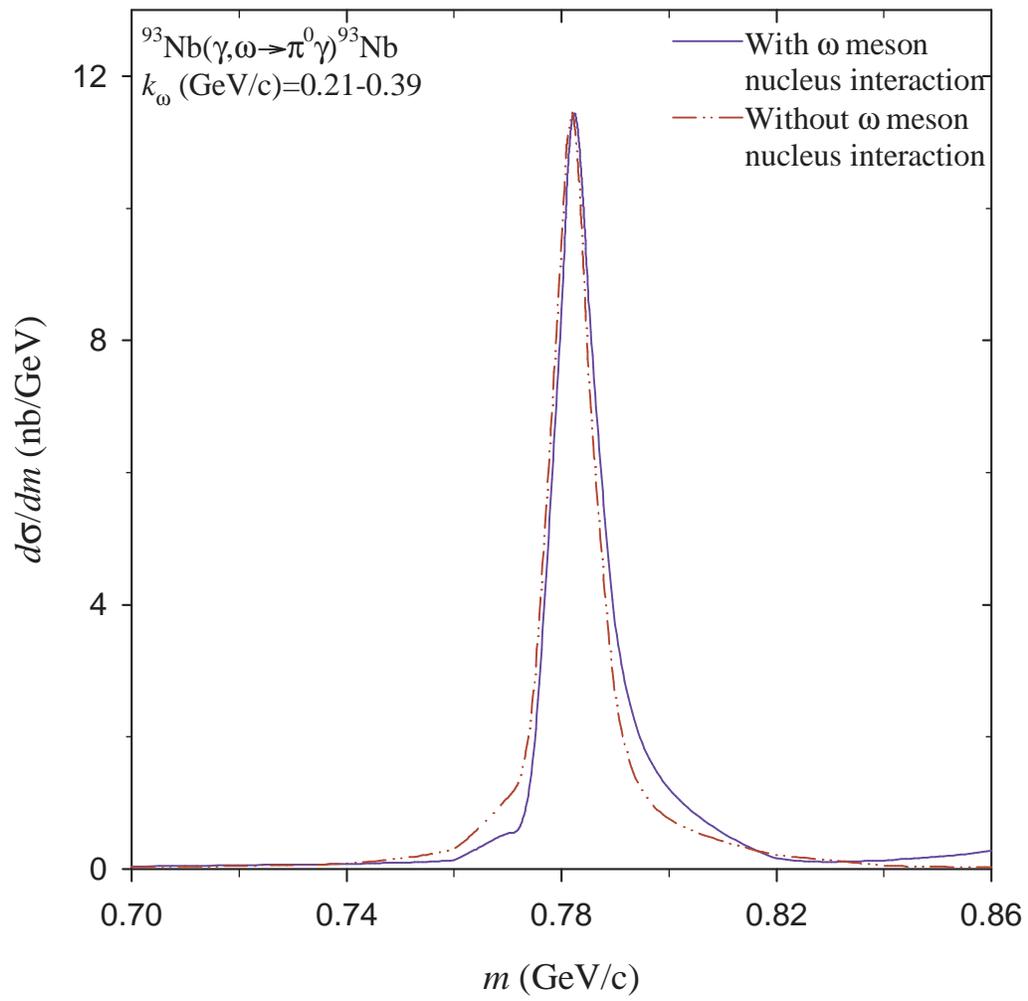
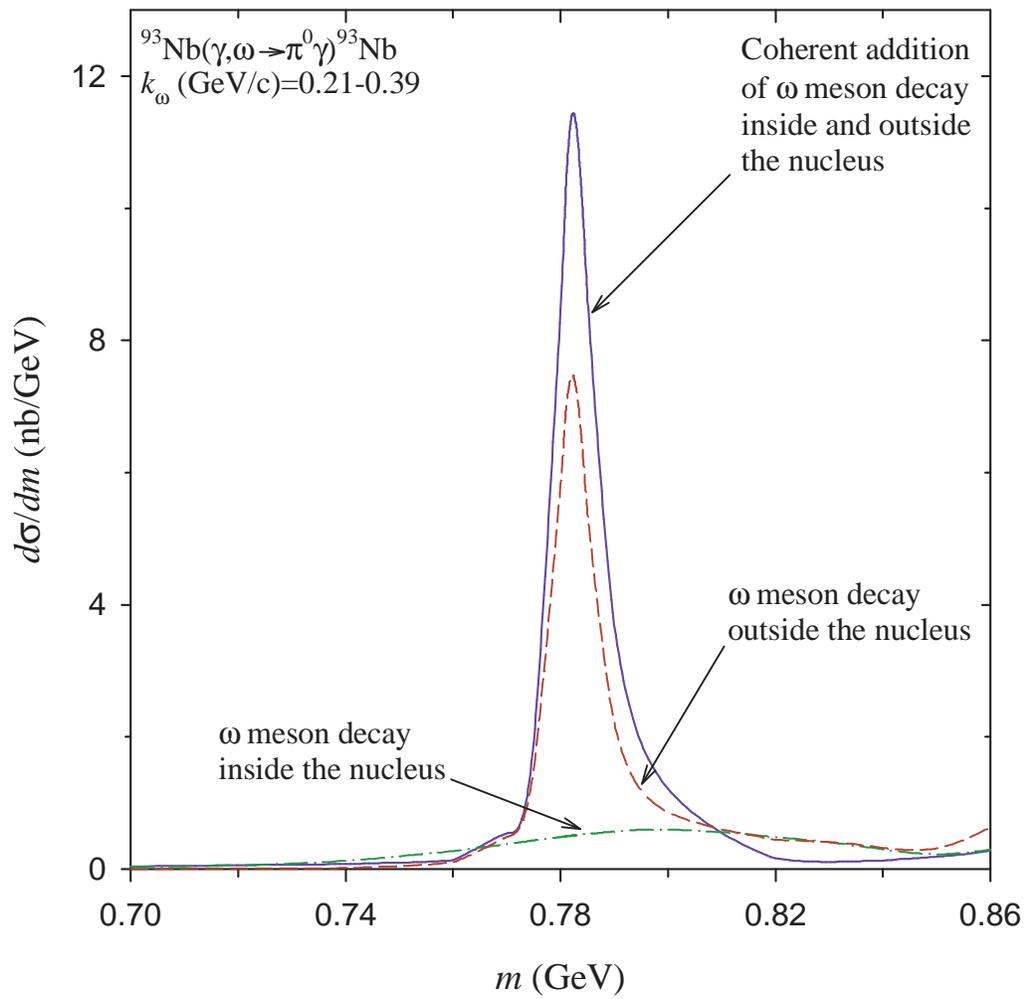
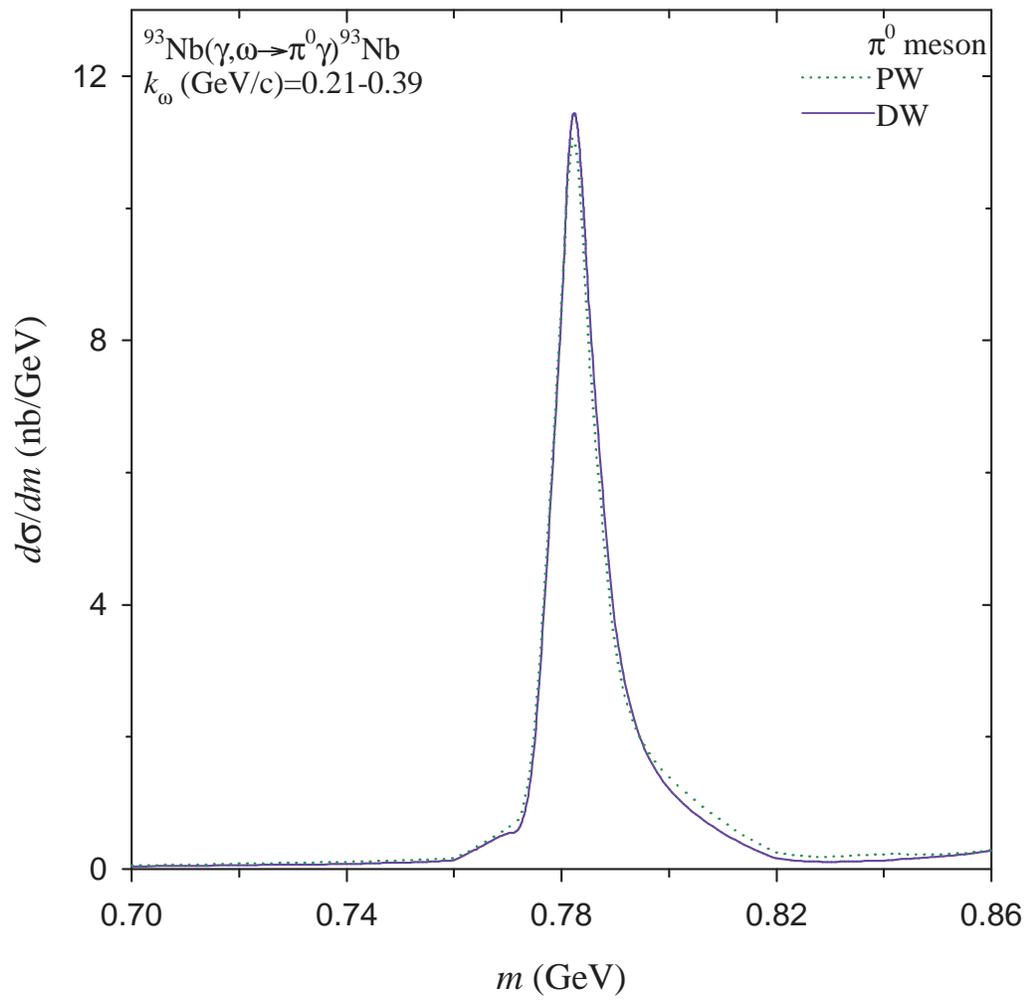
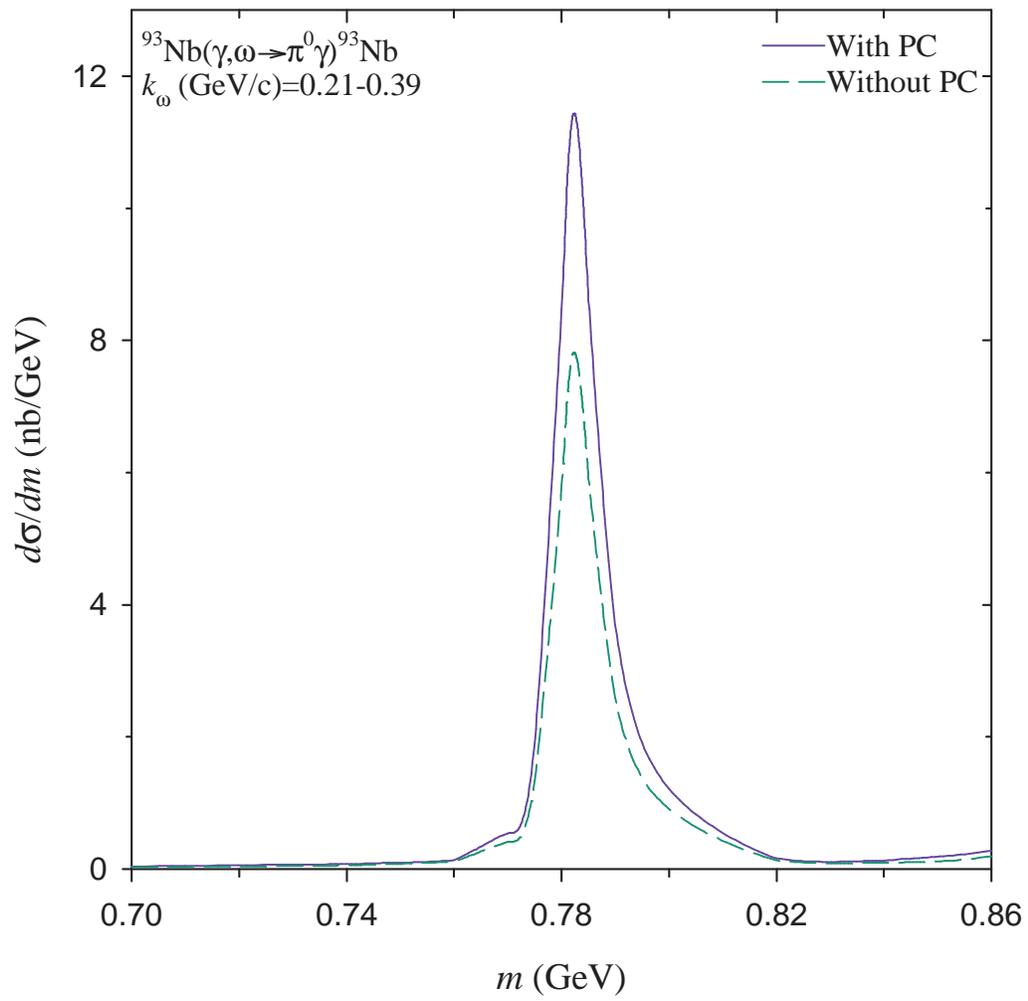
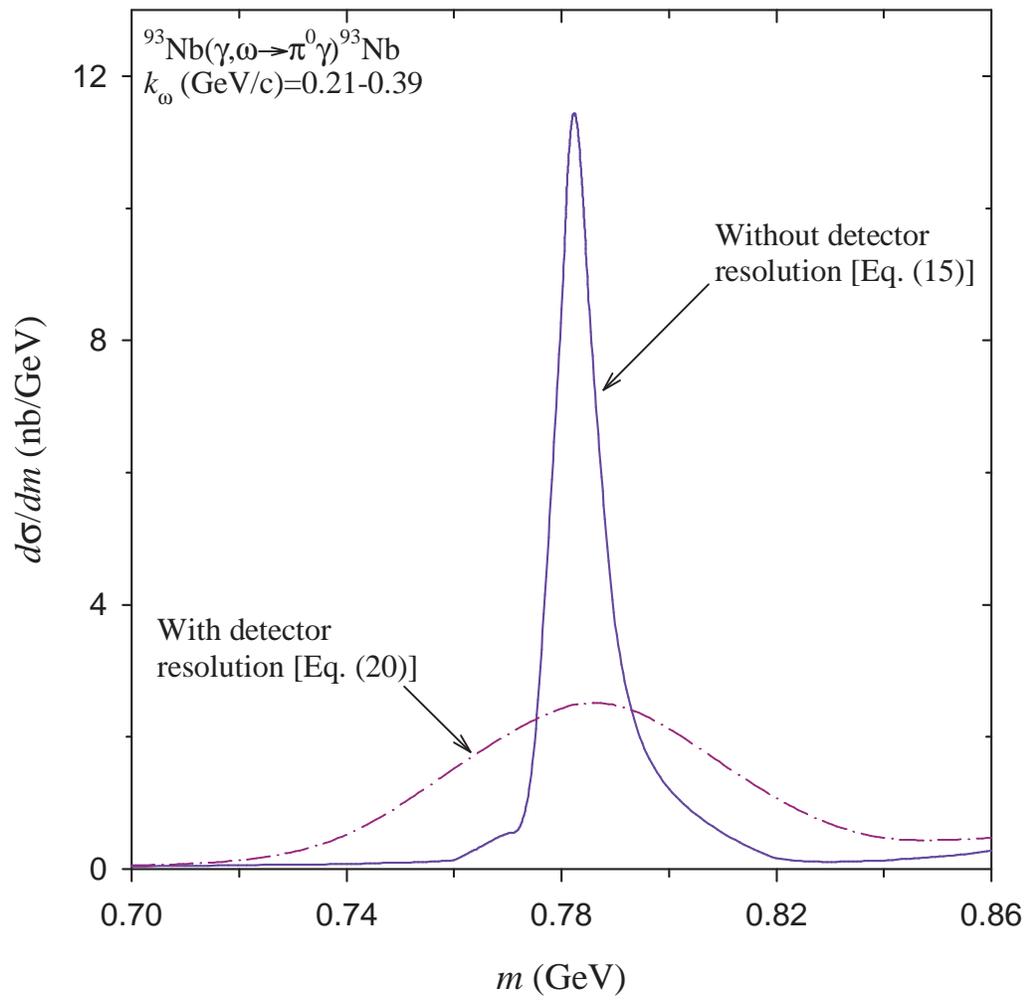
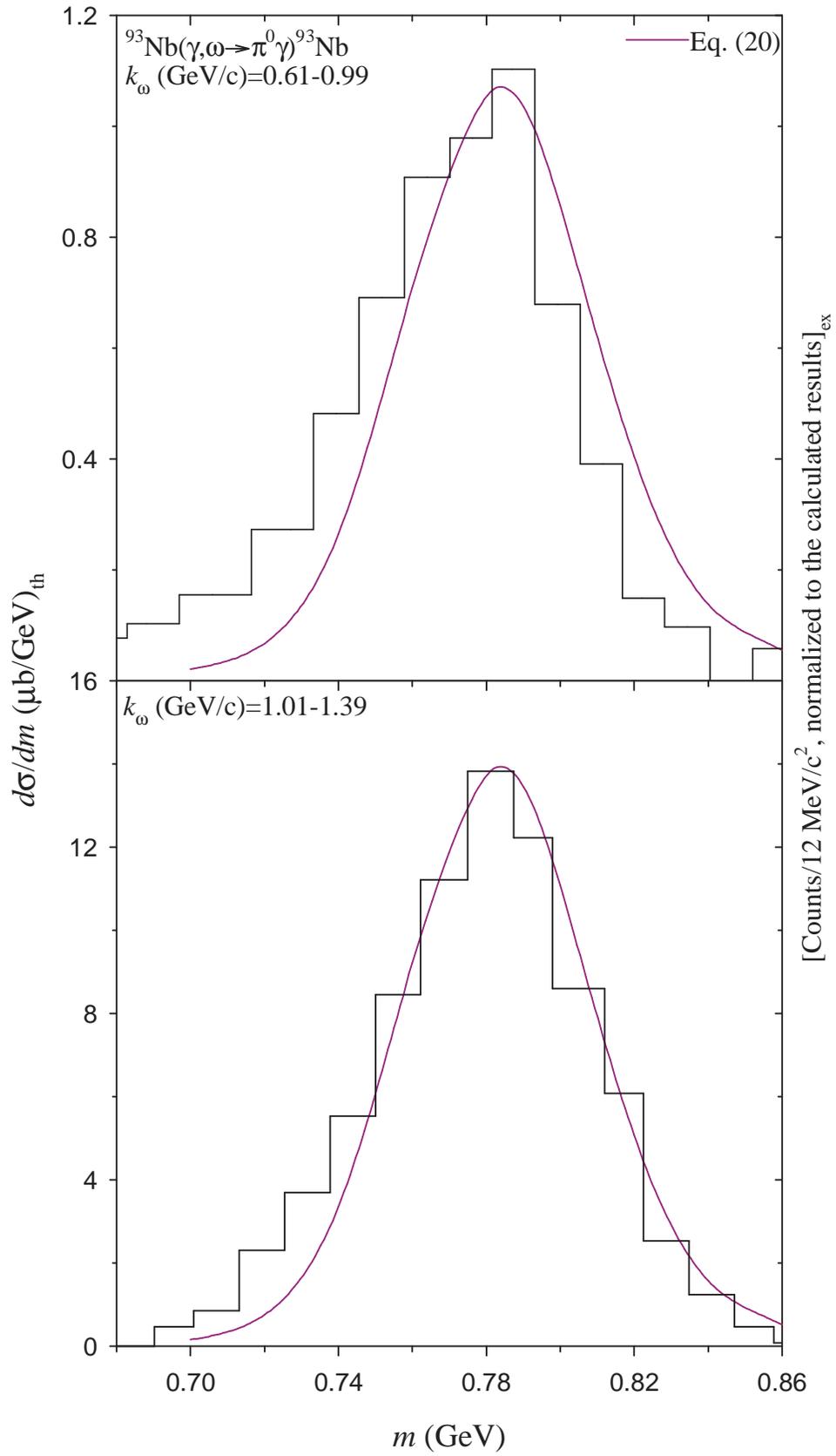
\end{document}